\documentclass[twocolumn]{revtex4}
\usepackage{graphicx}
\usepackage{epsfig}
\usepackage{subfigure}

\begin{document}

\title{Ellipsometry with an undetermined polarization state}

\author{Feng~Liu$^{1,2}$, Chris~J.~Lee$^{1}$, Juequan~Chen$^{1}$, Eric~Louis$^{1}$, Peter J.~M. van der Slot$^{2}$, Klaus J. Boller$^{2}$, and Fred~Bijkerk$^{1,2}$}
\affiliation{$^{1}$ FOM-Institute for Plasma Physics Rijnhuizen, Edisonbaan 14, 3439 MN Nieuwegein, The Netherlands \\
$^{2}$ Laser Physics and Nonlinear Optics Group, MESA+ Institute for Nanotechnology, PO Box 217, University of Twente, 7500 AE, Enschede, The Netherlands}

\begin{abstract}
We show that, under the right conditions, one can make highly accurate polarization-based measurements without knowing the absolute polarization state of the probing light field. It is shown that light, passed through a randomly varying birefringent material has a well-defined orbit on the Poincare sphere, which we term a generalized polarization state, that is preserved. Changes to the generalized polarization state can then be used in place of the absolute polarization states that make up the generalized state, to measure the change in polarization due to a sample under investigation. We illustrate the usefulness of this analysis approach by demonstrating fiber-based ellipsometry, where the polarization state of the probe light is unknown, and, yet, the ellipsometric angles of the investigated sample ($\Psi$ and $\Delta$) are obtained with an accuracy comparable to that of conventional ellipsometry instruments by measuring changes to the generalized polarization state.
\end{abstract}

\maketitle
\section{Introduction}
\label{sec:introduction}
Polarization-based measurements are increasingly important, both as a fundamental tool for scientific research~\cite{ISI:000290101000006,ISI:000290766600035,ISI:A1996VC33600031,ISI:000264285600035,ISI:000086119000045,ISI:A1991FN05600048,ISI:A1992JD95500021,ISI:A1997YJ86500045}, and as a vital tool for applications in the chemical, food, and pharmaceutical industries. Indeed, proposed classical communications systems, fluorescence measurements, non-invasive blood-sugar measurements, LIDAR, imaging through scattering media, strain, and temperature sensors all have implementations that rely on polarization based measurements~\cite{Gnauck:fba,Fu:2008ty,Dai:2010uh,Gaskell:2010p15633,Tsai:2008p2381}. In surface science, it is well known that polarization based measurements are very sensitive. For instance, ellipsometry has been used to measure sub-monolayer changes in surface coverage~\cite{Striebel:1994p261,VonKeudell:2002ua}. As such, it is a nearly-ideal technique for monitoring modern epitaxial fabrication techniques~\cite{Johs:1998wi} and for contamination monitoring of vacuum components (e.g., synchrotron optics)~\cite{Chen:2009p9263,Chen:2009vw}. The sensitivity of ellipsometry comes from the ability to set and measure polarization states with a high degree of accuracy. As a result, it is generally thought that, in order to make polarization-based measurements, one must know the polarization state of the probing light field throughout the optical train, and, especially, the polarization state of the light incident on the surface of interest. The corollary to this is that the use of optical components that disturb the polarization (e.g., optical fibers) require calibration, and, in the case of optical fibers, where temperature and stress induced birefringence have a large influence on the polarization state, this is not possible in all but the most limited circumstances. This is because, in contrast to what the name suggests, polarization maintaining (PM) fibers do not preserve arbitrary polarization states~\cite{Gisin:2002wz}. For instance, if linearly polarized light is injected at an arbitrary polarization orientation with respect to the axes of birefringence of the fiber, then the output is an elliptically polarized state with an unpredictable orientation and degree of ellipticity. As a consequence, fiber-based ellipsometry and polarimetry have relied on rather complicated experimental apparatus~\cite{Zhang:2003usa,Kim:2005jp}, use a wavelength that is only supported by one polarization mode in a PM fiber~\cite{Sun:2010vs}, or only provide qualitative information~\cite{YOSHINO:1984p11660}.

The core discovery that we present in this paper is that, given an input polarization state of light and an optical fiber that is subject to environmentally induced birefringence variations, the output state does not map to every point on the Poincare sphere, but rather, only a discrete set of polarization states are accessible. As the temperature, for instance, varies in time, the set of output polarization states appear as a fixed orbit on the surface of the Poincare sphere. In our work, we show that the existence of such orbits, as an intrinsic property of the fiber, can be seen as a higher dimensional type of polarization preservation that survives severe environmental perturbances. Once this orbit is known, polarization-based measurements can be made by measuring deviations from the orbit. This is very different from direct calibration, where an input polarization state is mapped to a single output polarization state for a known set of environmental parameters. Instead, a single input polarization state is mapped to a set of output states as the environment varies over some (generally unknown) range of temperatures and stresses. This approach turns a measurement problem---environmental noise---into an advantage by using the statistical properties of the noise to provide increased sensitivity.

In this paper, we present experimental data and modeling results that demonstrate and make use of the observation that the polarization follows a random path, but, importantly, remains bound to well-defined subspace. As an example of how such knowledge can be applied, we demonstrate a fiber-based ellipsometer. We show that this ellipsometer is capable of detecting a well-defined carbon layer that is less than 1~nm thick on top of a multilayer Bragg reflecting mirror.

\section{Experimental setup}
\label{sec:experimental}
\begin{figure}
\center
\includegraphics[width=7cm]{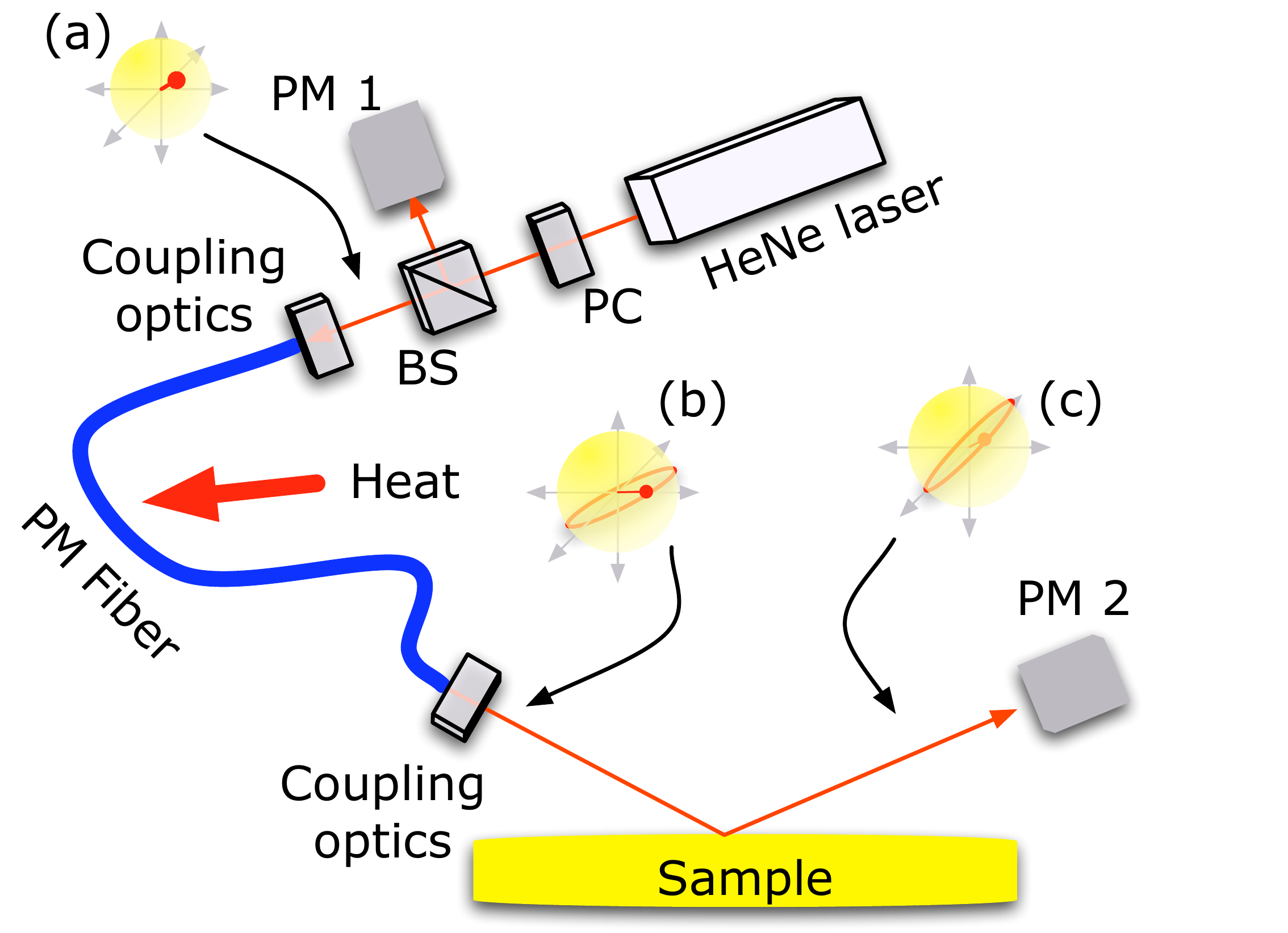}
\caption{Schematic of the ellipsometer setup. The polarization control optics (PC) set the light's input polarization, as illustrated on the Poincare sphere (a). After passing through a non-polarizing beam splitter (BS), part of the light is sent to a rotating compensator for analysis (PM 1) of the input polarization and part is coupled into the PM fiber. The polarization state of the light exiting the fiber falls somewhere on the orbit illustrated on Poincare sphere (b). The light reflects off the sample, rotating the orbit to that shown in Poincare sphere (c) and the final polarization state is measured at PM 2.}
\label{setup}
\end{figure}

The experimental setup is shown schematically in Fig.~\ref{setup}. As a light source, we used a Helium-Neon laser that emitted light that was nominally linearly polarized. The polarization of the light field was purified to a linearly polarized state, of which the azimuth angle was set to be parallel to the plane of the optical table, using a Wollaston prism. The Wollaston prism ensures that the polarization purity is 100000:1. The input polarization state to the fiber was controlled using a Soleil-Babinet compensator or a quarter-wave plate, which were manually controlled using precision rotation and translation stages. This allowed the individual relative Stokes vector intensities to be set within 0.02\% of a desired value. After setting the polarization, the laser beam was passed through a non-polarizing beam splitter, before being coupled into a 2~m long PM-fiber (Thorlabs, PM-630-HP) using a glass aspheric lens. Light was coupled out of the output end of the fiber and collimated with an identical coupling lens. The output fiber end was mounted on a manual rotation stage to set the angle of incidence with respect to the sample's plane of reflection. The accuracy of the angle of incidence on the sample was found to be 0.5$^{\circ }$. In all measurements, the middle section of the fiber (about 1.3~m) was coiled up on a metal cylinder and placed in a water bath with a heater to simulate environmental changes by varying the temperature between 19 and 30$^{\circ}$C. Only a small fraction of this range (2-3$^{\circ}$C) was used during the experiments, though, because that was sufficient to obtain the full range of accessible polarization states (see below).

After reflection from the sample, the polarization state of the light was measured using a rotating compensator, which consisted of a quarter waveplate, rotating at an angular frequency of 0.1 Hz, a Wollaston prism, and a Si PiN photodiode (PM 1 and 2 in Fig.~\ref{setup}). The photodiode voltage was observed on an oscilloscope and transferred to a computer for analysis. The photodiode signals were evaluated to obtain the polarization in terms of the Stokes vector components~\cite{Hauge:1975vw}. The polarization states are calculated as Stokes vectors and presented graphically in either 2D or 3D projections of the Stokes vector space. In order to restrict our analysis to the polarization properties of the PM fiber and sample, we compensate for the (small) modification of the polarization state due to the non-polarizing splitter with a predetermined matrix.

Ideally, neither the PM fiber nor the sample are changing, and the frequency components of the photodiode signal should only contain the DC component and two harmonics of the rotating compensator's angular frequency. However, this is true only when the polarization state is constant during the measurement period. In our case, the fiber's output states are constantly changing and, thus, allows a small non-zero amplitude for additional frequency components. By analyzing the amplitude of these components, the direct polarization state measurement noise was estimated to be 0.02$^{\circ }$ (for the ellipsometric parameter $\Delta$), which is comparable to a conventional Ellipsometer~\cite{Anonymous:2005una}.  

\section{Results}
\label{sec:Results}
\begin{figure}
\center
\subfigure[]{\includegraphics[width=6cm]{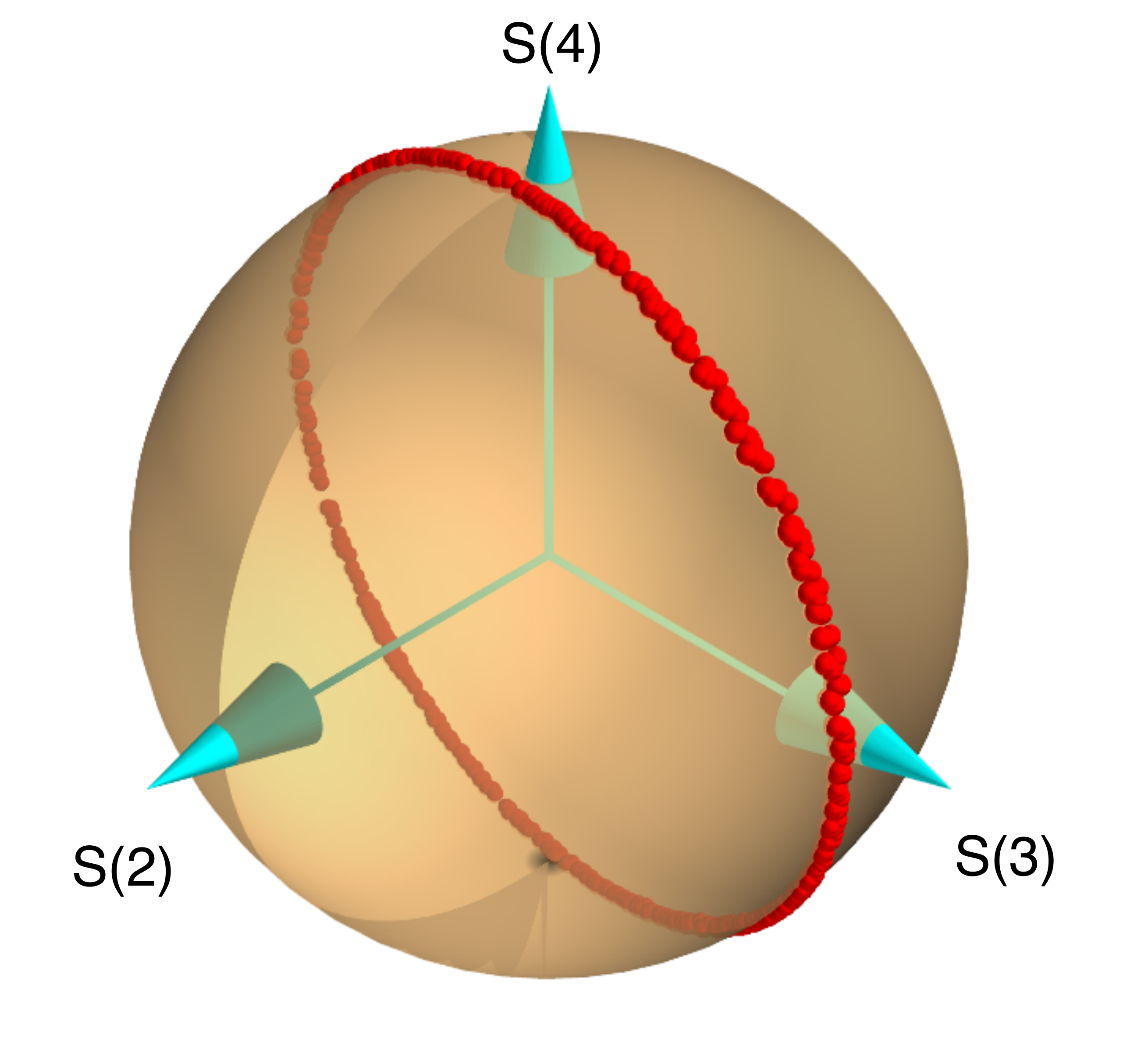}}\\
\subfigure[]{\includegraphics[width=6cm]{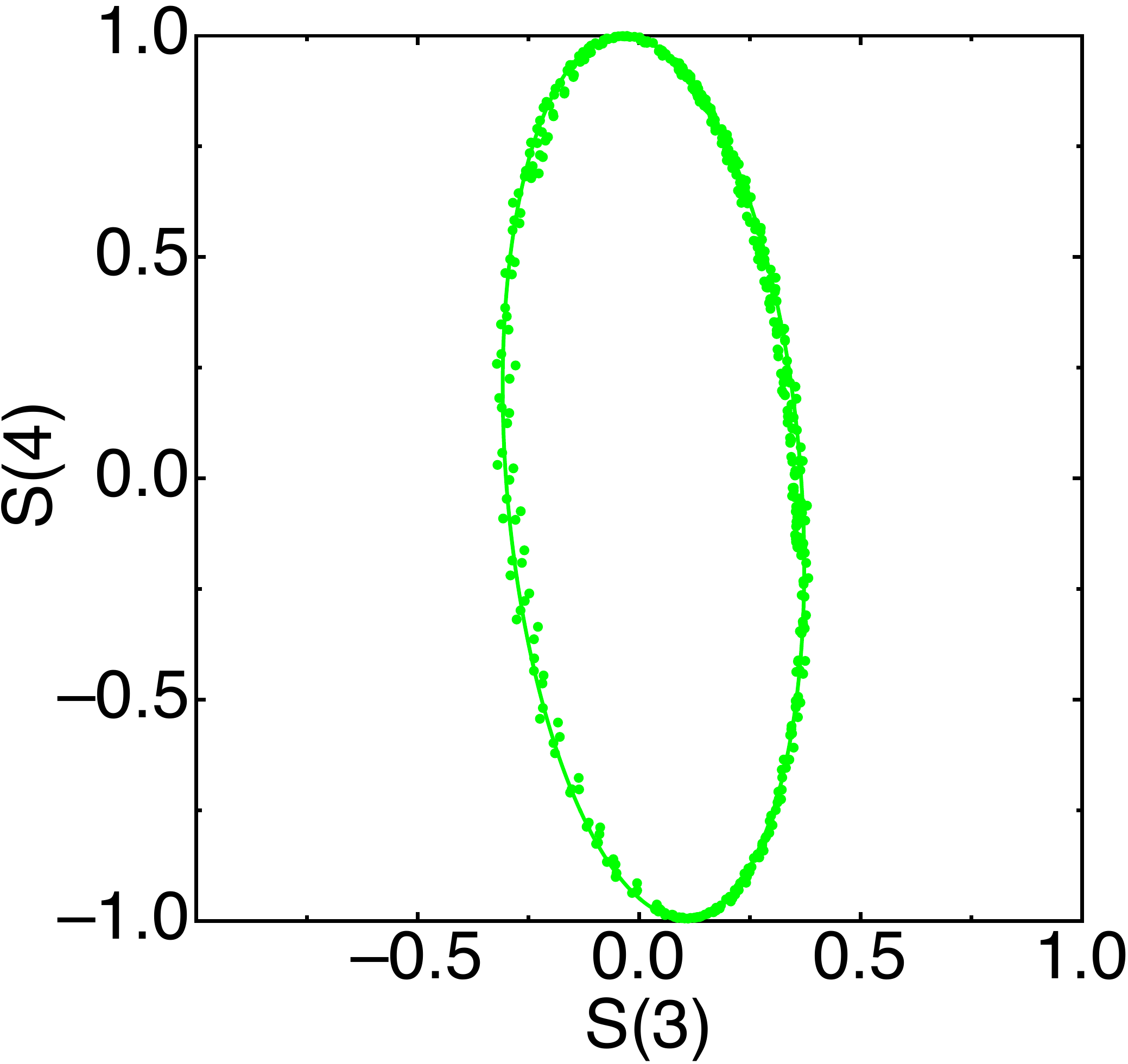}}
\caption{Polarization state of the light exiting the fiber for a fixed input polarization and varying the fiber temperature between 19.4 and 21.8$^{\circ}$C. Subfigure (a) shows a three dimensional image of the polarization state on the Poincare sphere, while (b) is the projection of that orbit onto the 3rd and 4th Stokes vector components. The solid line is a fit to the data with an ellipse (see below).}
\label{Polarization-sphere-exp}
\end{figure}

The measured polarization states of the output of the PM fiber are plotted in Fig.~\ref{Polarization-sphere-exp}(a) as zero-dimensional points on the two-dimensional surface of the Poincare sphere, where all data are, unless it is specifically stated otherwise, normalized so that the total intensity is unity. The notable feature is that despite the wide variation in temperature, the polarization states are not observed to fall randomly on the two-dimensional Poincare surface. Instead, a single, one-dimensional orbit is traced out, which is shown more clearly by taking a cross section through the sphere to obtain a projection of the orbit in a Stokes vector plane (Fig.~\ref{Polarization-sphere-exp}(b)). The analysis of the orientation of this orbit is key to performing polarization-based measurements without precise knowledge of the polarization of the probing light field.

To show that this polarization orbit can be used as a generalized, higher dimensional polarization state and be used to make polarization-based measurements, we show how the two ellipsometric parameters, $\Psi$ and $\Delta$ can be calculated from changes to the polarization orbit. We begin by considering light exiting a fiber in a set of polarization states denoted by $\mathcal{I}_{sp}^{xy}$ where the subscript and superscript denote the polarization coordinate systems of the sample ($s-p$) and the fiber ($x-y$). An isotropic reflecting surface transforms the polarization state of incident light according to the following Mueller matrix in the sample coordinate system~\cite{Anonymous:2005una}
\begin{equation}
M =A\cdot \left[ 
\begin{array}{cccc}
1 & \cos 2\Psi  & 0 & 0 \\ 
-\cos 2\Psi  & 1 & 0 & 0 \\ 
0 & 0 & \sin 2\Psi \cos \Delta  & \sin 2\Psi \sin \Delta  \\ 
0 & 0 & -\sin 2\Psi \sin \Delta  & \sin 2\Psi \cos \Delta 
\end{array}%
\right]   \label{iso refle surf}
\end{equation}
where $A = (\left\vert r_{x}\right\vert ^{2}+\left\vert r_{y}\right\vert ^{2})/2$ and $\left\vert r_{x}\right\vert ^{2}$ and $\left\vert r_{y}\right\vert ^{2}$ are intensity reflectivities for the $P$ and $S$ polarized light respectively. The initial states become
\begin{equation}
\label{eq:transform}
\Re^{xy}=M \cdot \mathcal{I}_{sp}^{xy}
\end{equation}
In other words, the ellipsometric information of the sample is carried in a global transformation from $\mathcal{I}_{sp}^{xy}$ to $\mathcal{R}^{xy}.$

The matrix $M $ is a direct sum of the matrices of its two block diagonal subspaces. The (desired) values of $\Psi $ and $\Delta $ can, thus, be obtained separately.

The set of polarization states of the light field incident on, and reflected by the sample are given by
\begin{equation}
\label{eq:stokes}
\begin{array}{cc}
	\mathcal{I}_{sp}^{xy} = \left[\begin{array}{c}
		I_{1}\\
		I_{2}\\
		I_{3}\\
		I_{4}
		\end{array}\right],
	&
	\mathcal{R}_{xy} = \left[\begin{array}{c}
		R_{1}\\
		R_{2}\\
		R_{3}\\
		R_{4}
		\end{array}\right]
\end{array}
\end{equation}
substituting equations~\ref{eq:stokes} into equation~\ref{eq:transform} gives
\begin{eqnarray}
\label{eq:RforPsi1}
R_{1} &=& A(I_{1}-I_{2}\cos 2\Psi) \\
\label{eq:RforPsi2}
R_{2} &=& A(I _{2}-I_{1}\cos 2\Psi)\\
\label{eq:RforDelta1}
R_{3} &=& A\sin(2\Psi)(I_{3}\cos\Delta - I_{4}\sin\Delta)\\
\label{eq:RforDelta2}
R_{4} &=& A\sin(2\Psi)(I_{3}\sin\Delta + I_{4}\cos\Delta)
\end{eqnarray}
Dividing Eqs. \ref{eq:RforPsi1} by \ref{eq:RforPsi2} and rearranging gives:
\begin{equation}
\label{eq:Psi}
	\Psi = \frac{1}{2}\arccos\left(\frac{R_{r}I_{r} - 1}{R_{r} - I_{r}}\right)
\end{equation}
where $R_{r}= R_{2}/R_{1}$ and $I_{r}= I_{2}/I_{1}$. $\Delta$ is obtained by requiring consistency between equations~\ref{eq:RforDelta1} and \ref{eq:RforDelta2}. Note that equations~\ref{eq:RforDelta1} and \ref{eq:RforDelta2} have the form of a scaling factor and a rotation, which, when applied to an ellipse, reduces its area and changes the orientation of its major axis in Stokes space. In practice, since the polarization orbit, when projected in the plane of the 3rd and 4th Stokes vectors, is an ellipse, $\Delta$ is the angle difference between the orientations of the ellipses given by $\mathcal{I}_{sp}^{xy}$ and $\mathcal{R}^{xy}$.
 
It is important to note that our analysis assumes that all states are measured in the $x-y$ coordinate system. Slight changes of alignment between the coordinate system of the fiber and the sample introduces an extra rotation to the polarization state that depends on the misalignment angle, $ \delta \alpha$. In that case, it can be shown that the systematic error added to $\Psi$ and $\Delta$ is of the order of $\delta\alpha^{2}$. Provided that the physical alignment of the PM-fiber is held sufficiently constant, the uncertainty in $\Delta$  and $\Psi$ should be comparable to that of conventional ellipsometry.

\begin{figure}
\center
\subfigure[]{\includegraphics[width=6cm]{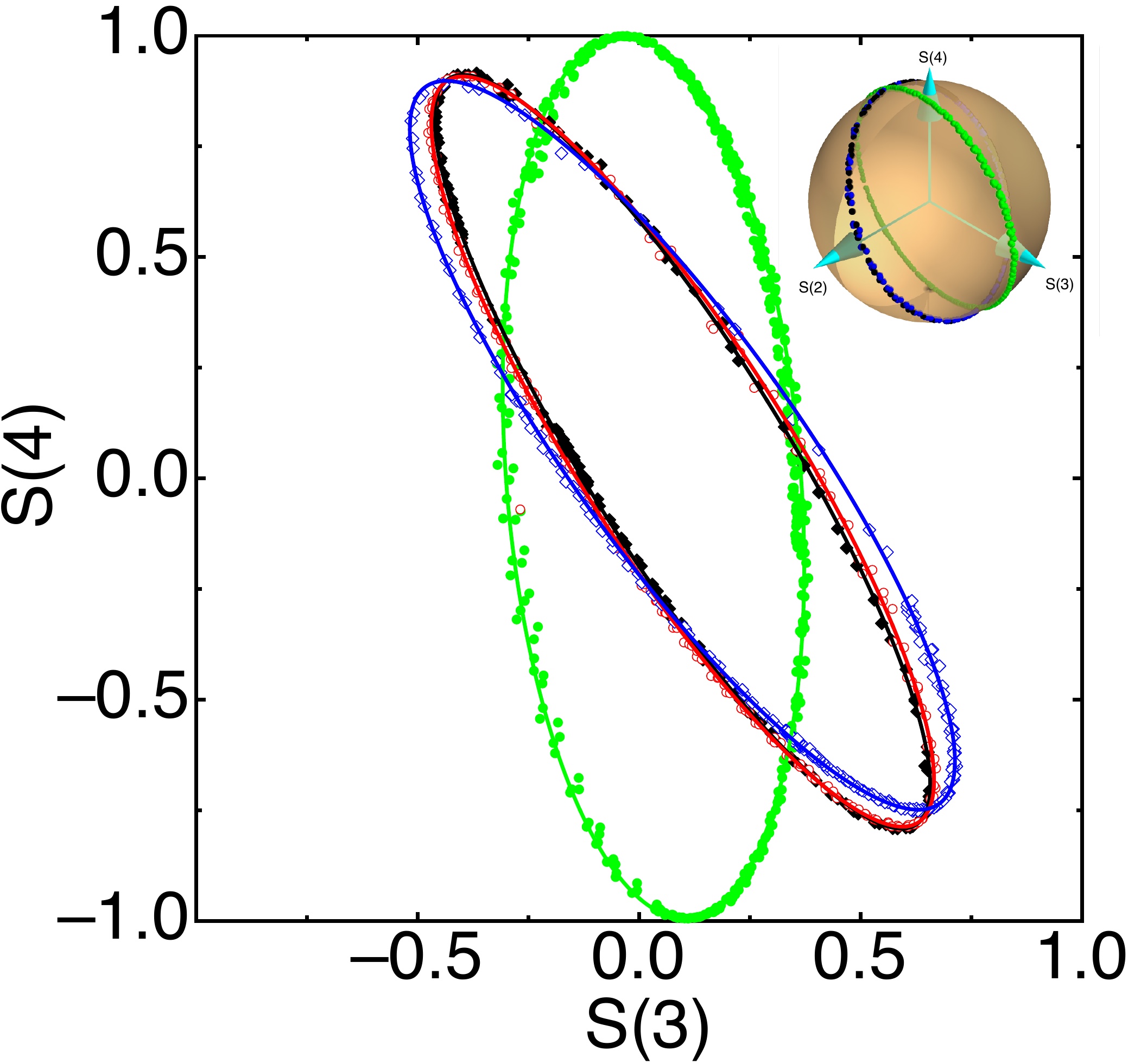}}\\
\subfigure[]{\includegraphics[width=6cm]{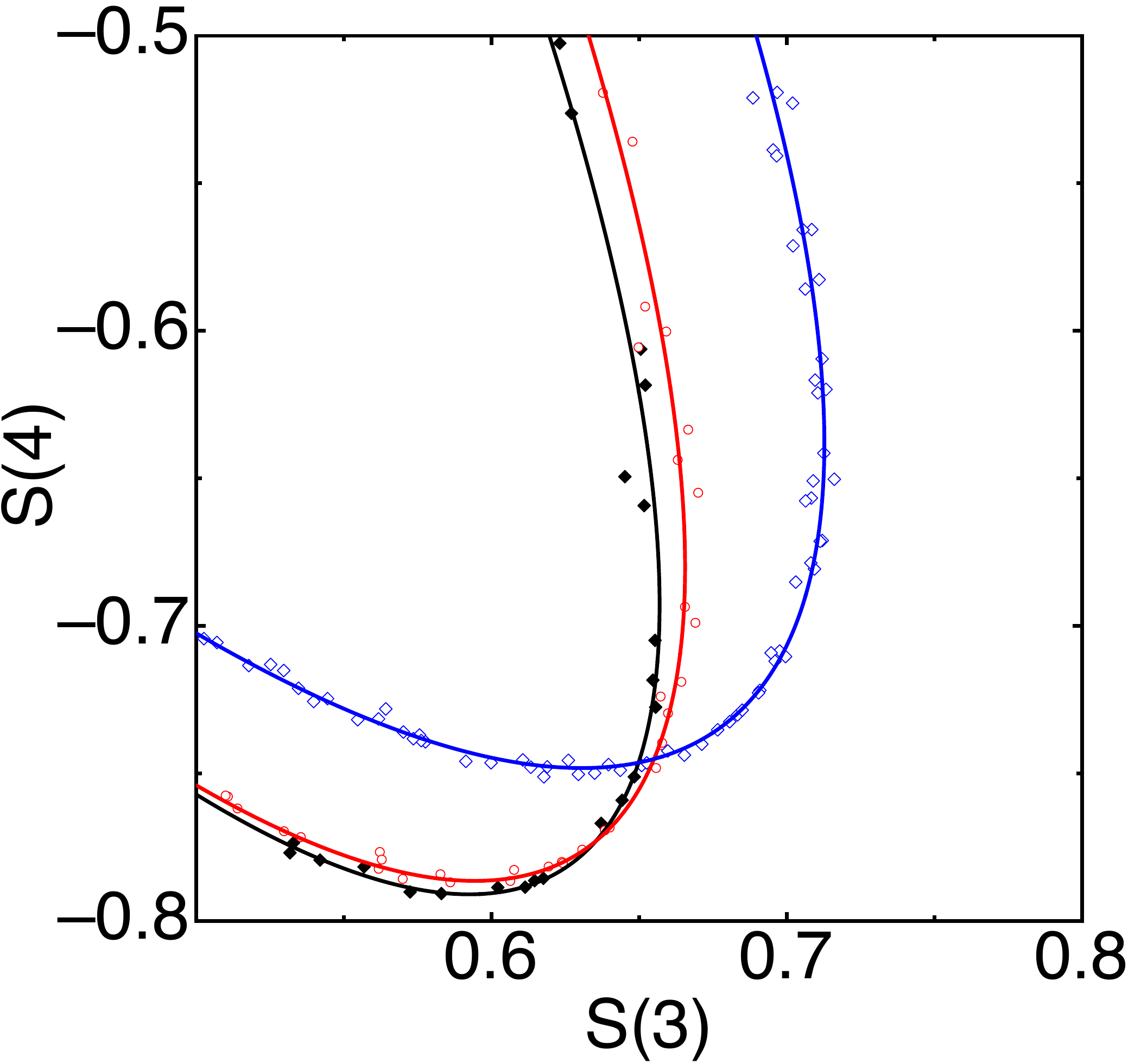}}
\caption{Subfigure (a): projection of the polarization orbits on the plane of the third and fourth Stokes vector components for the light exiting the fiber (solid green circles), after reflection from: a multilayer mirror (filled black diamonds), a multilayer mirror with a 0.3~nm thick carbon layer (open red circles), and a multilayer mirror with a 0.8~nm thick carbon layer (open blue diamonds). Subfigure (b) shows a zoomed section of the projection of the polarization orbits. The lines are the fitted ellipses, which are used to determine $\Psi$ and $\Delta$.}
\label{carbon samples}
\end{figure}

To demonstrate the applicability of this analysis, we performed fiber-based ellipsometry on three multilayer Bragg reflecting samples~\cite{Louis:2011kka}, two of which have been coated with amorphous hydrogenated carbon~\cite{Chen:2009p9263,Chen:2009vw}. Spectroscopic ellipsometry (Wollam M2000) revealed that the carbon layers on these samples are 0.3, and 0.8~nm thick, while the third, supposedly uncoated sample, has $\sim$0.1~nm naturally occurring carbon layer with a different composition to that of the first two samples. Both conventional and fiber-based ellipsometry measurements were performed at an angle of incidence of 66$^{\circ}$. The raw ellipsometric data from the fiber-based ellipsometer, using the experimental procedure described described above, is shown in Fig.~\ref{carbon samples}(a). The measured orbits have been projected onto the plane of the third and fourth Stokes vector components (the inset shows the three dimensional representation of the data). As illustrated by the zoomed in section, shown in Fig.~\ref{carbon samples}(b), the data for each sample are systematically modified by the sub-nanometer layers of carbon to generate three different ellipses. The data for each sample show very little deviation from the fitted ellipse.

Using the analysis described above, we obtain the $\Psi$ and $\Delta$ values presented in Tab.~\ref{Psi and Delta values}. These values agree well with those obtained from a conventional ellipsometer. The small systematic difference between the values from the two instruments is due to a small difference in the angle of incidence between the two measurements (the accuracy of the angle of incidence is $\pm$0.5$^{\circ}$).

\begin{table}[htdp]
\caption{$\Psi$ and $\Delta$ values for fiber-based ellipsometric measurements (columns 2 and 3) and conventional ellipsometer measurements (columns 4 and 5) on carbon-coated MLMs}
\begin{center}
\begin{tabular}{|c|c|c|c|c|}
\hline
Sample & $\Psi$ $^{\circ}$ & $\Delta$ $^{\circ}$ & $\Psi$ $^{\circ}$ & $\Delta$ $^{\circ}$ \\
\hline
Bare MLM & 28.04 & 152.5 & 28.158 & 152.77\\
\hline
0.3~nm & 28.00 & 152.1 & 28.265 & 152.39\\
\hline
0.8~nm & 28.48  & 148.9 & 28.389 & 150.81\\
\hline
\end{tabular}
\end{center}
\label{Psi and Delta values}
\end{table}%

\section{Discussion}
\label{sec:discussion}
To understand why PM fibers that are subject to varying temperature and strain, produce a polarization orbit, a numerical model of a non-ideal PM fiber was developed.  In simple terms, the fiber is a strongly birefringent material with a time and space-dependent beat length, and some degree of polarization mode cross-talk. In this model, the orientation of the axes of birefringence and the refractive index difference between the two axes changes as a function of position along the fiber and time. To analyze the effect of such a fiber, we break the fiber up into ideal segments, $\Delta l$. Mathematically, this corresponds to rotating the coordinate system of the fiber for light entering a fiber segment and then reversing the rotation upon exiting. The phase delay experienced by the light field then depends on its orientation relative to the rotated coordinate system. This introduces a polarization change that is purely due to the relative phase delay of the components of the electric field along the local axes of birefringence.  The transfer matrix of the entire fiber is then given by a series of matrix multiplications~\cite{Anonymous:2005una}:
\begin{equation}
M_{fc}=\prod\limits_{l=0}^{L}R(\Delta \alpha (l, t))m_{f}(\frac{\Delta l}{b(l, t)}%
)R(-\Delta \alpha (l,t))  \label{fiber cross talk}
\end{equation}
$R(\Delta\alpha(l, t))$ and $R(-\Delta\alpha(l, t))$ are rotations due to an angular ($\Delta\alpha$) misalignment of the axes of birefringence of two neighboring ideal segments, while $m_{f}$ is the matrix transfer function due to the phase delay. The phase delay is expressed as a function of the beat length, $b(l, t)$, normalized to the segment length. Both $b(l, t)$ and $\Delta\alpha(l, t)$ are time dependent in a constrained but random manner. In these simulations, the maximum limits of $\Delta\alpha$ and changes to $b(l,t)$ were given by the yield strength of glass---the strain required to introduce these variations should not break the fiber. To allow direct comparison between the model results and experimental results, the average variation in $\Delta\alpha$ was determined by specifying a maximum polarization cross-talk per unit length. 

\begin{figure}
\center
\includegraphics[width=6cm]{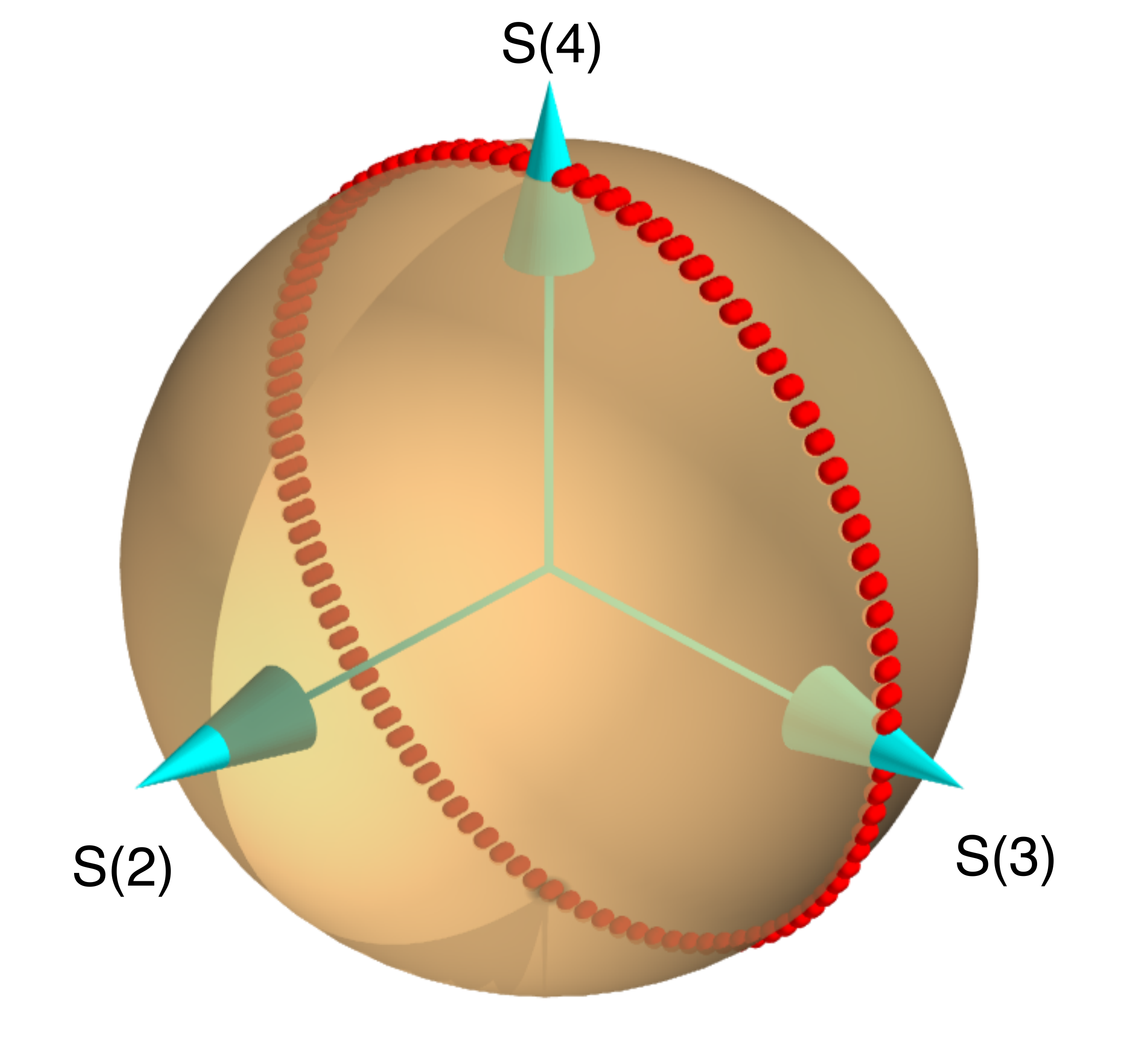}
\caption{Numerical evaluation of equation \ref{fiber cross talk} for a total polarization mode cross talk of 20 dB. Time-dependent temperature changes cause changes to the local degree and orientation of the fiber's birefringence. The resulting output is an orbit on the Poincare sphere.}
\label{model-results}
\end{figure}

As the phase change varies as a function of position and time in the fiber, the output polarization states form a simple, circular orbit in Stokes space. The effect of cross-talk is to introduce deviations from a simple orbit in Stokes space, instead, a more complicated, but repeating, orbit is obtained. Numerically evaluating equation \ref{fiber cross talk} under conditions that the maximum total cross talk is 20~dB, which is the maximum specified cross-talk for our fiber, shows (see Fig.~\ref{model-results}) that, even in the presence of a small amount of cross-talk, the polarization orbit can still be approximated by a simple orbit (the elongation of the red markers in Fig.~\ref{model-results} indicate the deviation from a simple orbit). Importantly, these simulations indicate that, as long as we can quantify the polarization orbit of an optical component, then it will be possible to use light exiting that component for polarization-based measurements. It should be noted there is the (slim) possibility that the polarization change induced by the sample is a rotation in the plane of the polarization orbit. In this case, the sample is undetectable. This problem can be avoided by making measurements at more than one angle of incidence.

In conclusion, we have demonstrated that sensitive polarization-based measurements are possible even when the polarization state of the probing light field has been passed through an optical element with an unpredictably fluctuating birefringence. This is possible because, for a fixed input polarization, the environmentally induced fluctuations in the output polarization after propagation through an optical element or material still lie in a single orbit of the Poincare sphere, creating what we have termed a generalized, higher dimensional polarization state. Polarization-based measurements can then be made by analyzing changes to the higher dimensional state, rather than the individual polarization states that make up the higher dimensional state. To illustrate this, we demonstrated a fiber-based ellipsometer, capable of detecting carbon layers with a thickness of 0.3~nm.

\section{Acknowledgements}
FL, CJL, and FB acknowledge funding from M2i and ``Controlling photon and plasma induced processes at EUV optical surfaces (CP3E)'' of the ``Stichting voor Fundamenteel Onderzoek der Materie (FOM)'', which is financially supported by the ``Nederlandse Organisatie voor Wetenschappelijk Onderzoek (NWO)''. The CP3E programme is co-financed by Carl Zeiss SMT and ASML.


\end{document}